# Efficient Context Management and Personalized User Recommendations in a Smart Social TV environment

Fotis Aisopos[i], Angelos Valsamis[ii], Alexandros Psychas[i], Andreas Menychtas[i] and Theodora Varvarigou[i]

[i]Distributed, Knowledge and Media Systems Group, National Technical University of Athens, Greece
*{fotais, alps, ameny }@mail.ntua.gr, dora@telecom.ntua.gr*
[ii]Department of Informatics and Telecommunications, National and Kapodistrian University of Athens, Greece
*ang.valsamis@gmail.com*

**Abstract.** With the emergence of Smart TV and related interconnected devices, second screen solutions have rapidly appeared to provide more content for end-users and enrich their TV experience. Given the various data and sources involved - videos, actors, social media and online databases- the aforementioned market poses great challenges concerning user context management and sophisticated recommendations that can be addressed to the end-users. This paper presents an innovative Context Management model and a related first and second screen recommendation service, based on a user-item graph analysis as well as collaborative filtering techniques in the context of a Dynamic Social & Media Content Syndication (SAM) platform. The model evaluation provided is based on datasets collected online, presenting a comparative analysis concerning efficiency and effectiveness of the current approach, and illustrating its added value.

## 1. Introduction

Recent studies show that mobile devices are gradually employed more and more in parallel with TV usage, creating the so called second screen phenomenon[1]. Users comment or rate TV shows on Social Media and search for related information about actors, places and all other sorts of information related to the show they are watching. This phenomenon is expected to grow exponentially, creating a huge impact on the way content is created and delivered, not only through regular broadcasting but also thought mobile devices. However, there are still no second screen standards, protocols or ever common practices for users to discover and access additional information related to the consumed content [1].

The lack of such attributes drives them to search intuitively in the social media or online search engines for content. Moreover, despite the popularity of products like Vizrt Social TV [2] or Beamly social content network, building a custom user experience, including context management and recommendations, is inefficient and not scalable[3]. The continuing growth of online content though led to a need for the creation of systems that manage it, to provide quality of service, content syndication and recommendations. The main motivation of the current work is to provide an efficient model that captures and manages Social TV-related context data, to provide smart recommendations to the users.

---

[1] Second Screen Society: http://www.2ndscreensociety.com/

[2] http://www.vizrt.com/solutions/social-tv-solution/

[3] http://adexchanger.com/digital-tv/social-tv-platform-beamly-learns-the-second-screen-is-a-feed/

In the context of the SAM Project [2], a Social Media-aware content delivery platform for syndicated data to be consumed in a contextualised social way through second screen devices has been provided. The former, out-dated model of users searching for the information they desire is replaced with a new approach, where information reaches users on their second screen using content syndication. This paper focuses on user interaction history with first and second screen and the related behavioural models applied in SAM, to form an innovative Context Management mechanism, based on a graph database. To this end, users, Social Media items (e.g. widgets appearing in second screen) as well as related interactions are saved in the form of a nodes/edges, where graph analysis models and correlation techniques are employed to properly assess the relevance of each media item to every user. As a result, a relevance rating list for each first/second screen user is produced, allowing personalized recommendations of videos and related assets. Thus, user's Social TV experience is enriched as such: upon interaction with first screen, a relevant list of videos is recommended to her, while upon watching a specific video only relevant information widgets (like wikipedia articles) appear in second screen.

More specifically, the current paper contributions are summarized below:

- Scalable and timely efficient Social Media user profiling and Context Management using an appropriate graph model (visualizing users, assets, interactions)
- Intelligent data analysis based on a combination of graph analysis and collaborative filtering
- Multi-level Social TV personalized user recommendations via relevance rating approach for root assets (videos) in first screen and sub-assets (video-related sources of information) in second screen
- A comparative analysis of commonly used machine learning algorithms and clustering techniques, applied to Social TV user context-based recommendations

The rest of the document is organized as follows: Section 1 is the current one and serves the purpose of the introduction. Section 2 presents related work on graph-based and collaborative filtering techniques, investigating the existing Social TV user context modelling and recommendations. Section 3 introduces the SAM context management approach, while Section 4 elaborates on graph analysis and collaborative filtering. Finally, Section 5 analyzes the experimental results and Section 6 presents the conclusions.

## 2. Related Work

Graph databases have been extensively used lately to optimize storage and processing of highly connected data. For example, authors in [3],[4],[5] provide insights into Neo4j graph database and its performance advantages, illustrating the various cases it is used, including recommender systems that apply "item-to-item" and "user-to-user" (i.e. collaborative filtering) correlation. Demovic et al. [6] presented an interesting context-based graph recommendation approach, saving multimedia-related data in a graph structure and using Graph Traversal Algorithms to efficiently address user preferences. This work uses explicit user "likes" for movies or genres, but does not collect any other contextual or social data. Focusing on Social TV Platforms, works in [7] and [8] highlight the concept of context management and analysis in the frame of social enabled content delivery to multi-screen devices. These papers present a novel solution for media content delivery, based on the idea of fusing second screen and content syndication, exploiting the advancements in the area of social media.

Recommender systems for eCommerce [9],[10] usually follow a personalised recommendation approach, based on users' clustering and correlation, as well as behavior factorization [11]. When it comes to TV-related personalised recommendations, a substantial amount of work has been performed, focusing on TV programs and movies' context evaluation. Krauss et al. [12] introduced personalized TV program recommendations based on users' viewing behavior and ratings, combining various data mining approaches. A ten-fold cross validation over a user-generated dataset aggregated from the operation of the TV Predictor software resulted into a promising program prediction accuracy. Kim et al. [13] on the other hand, presented an automatic recommendation scheme based on a user clustering approach that did not require explicit ratings from TV viewers, but rather the watching history logs. The proposed rank model used a collaborative filtering technique, taking into account the watching times, to illustrate effectiveness with rich experimental results over a real usage history dataset.

Collaborative filtering techniques are frequently used by online recommender systems [14] in domains such as web services [15], social networks [16] or movies [17] selection. Kwon and Hong [18] propose a personalized program recommender for smart TVs using memory-based collaborative filtering, employing a novel similarity method that is robust to cold-start conditions and faster than existing approaches. The evaluation uses an own-built crawling agent to retrieve movie reviews by real users and predict ratings for non-viewed programs. On the other hand, work in [19] proposes improvements to two of the most popular approaches to Collaborative Filtering, introducing a new neighborhood based model, as well as extensions to SVD-based latent factor models and integrating implicit feedback into the model. Those are evaluated with a very limited form of implicit feedback, available on Netflix. Efficient methods for collaborative filtering like Item-to-Item or SVD [20] decrease the impact of noise and improve the ability for high quality recommendations systems, such as movie recommenders [21]. However in our case, the high performance of the graph analysis that will be presented is supported by a Pearson correlation technique, only in the case the user has not interacted with neighboring multimedia items, in which case Item-to-Item approaches would not provide sufficient results.

SAM Context Management for Smart Social TVs attempts to progress beyond the state-of-the-art solutions presented above, by providing a personalized multi-level recommendation mechanism (applying both for first and second screen content), based on an efficient graph-based approach. The localization of the graph analysis, in contrast with the global machine learning or collaborative filtering models, yields a high scalability for a big datasets of multimedia items and interactions.

## 3.  SAM Context Management Approach

### 3.1. Platform Architecture and Data collected

SAM aims at the development of a context-centric middleware that acts supportively to its advanced federated social media delivery platform, providing open and standardised way of defining, characterising, discovering, syndicating and consuming media assets interactively. In the context of SAM Platform, the generic components of Context Control are responsible for storing and managing context information. Figure 1 [22] shows the subcomponents realising the Context Representation operations along with the connections between them and other SAM components. The core component for context-related operations is the Context Manager, collecting contextual information from Social Media, including SAM dynamic communities exposed by Community Structure Analyser, as well as the Syndicator and the Dashboard. Based on the analysis of those data, the Context Manager produces ranked lists of assets (videos or widgets) per user and forwards those to the Syndicator that uses them to send smart recommendations to first and second screen, after a user logs in to the SAM Dashboard.

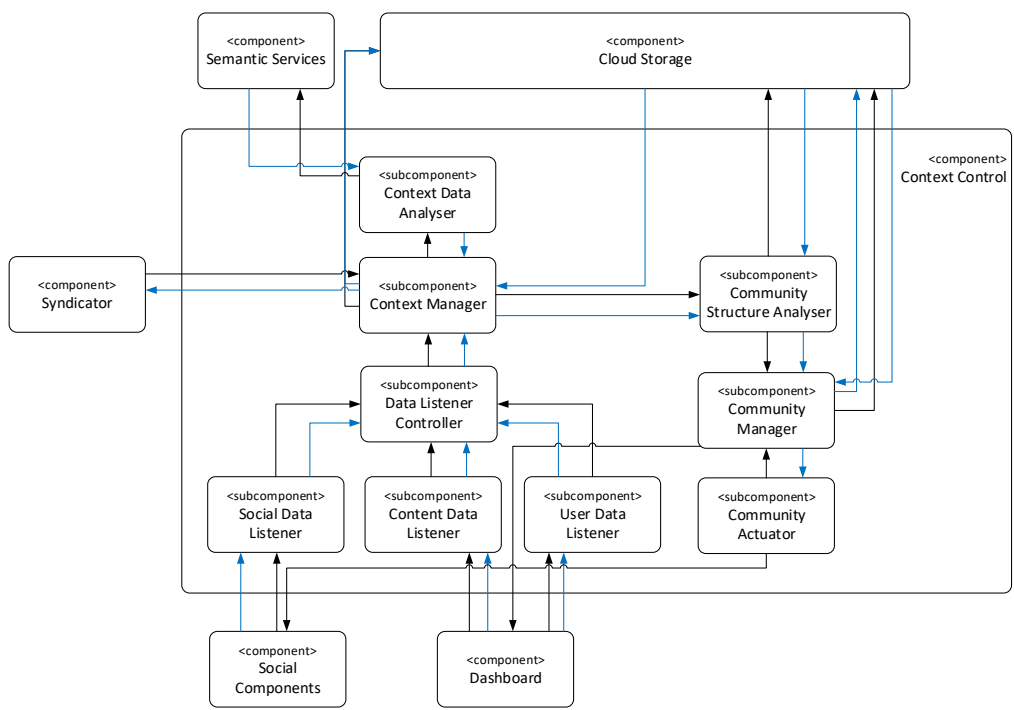

**Figure 1:** SAM Context Control Components

The SAM Dashboard connects to specific Data Listeners capturing and storing all interactions according to an extended W3C Social Web Working Group context model (successor to the OpenSocial format).The aforementioned Data Listeners (Social Data Listener, Content Data Listener, User Data Listener) are managed by a Data Listener Controller and forward to the Context Manager all data related to the Social TV context, including Social Media posts on SAM multimedia and user interactions with first and second screen. The interactions which are useful for the Context Data Analyser are the ones illustrating user's relevance or satisfaction with the content provided, such as "likes" or "scrolls", as well as text comments or online posts, further evaluated with the support of Semantic Services for sentiment analysis, so as to enrich the user profile with contextual information.

In specific, the following user interaction items are collected from Generic Dashboard listeners and sent to SAM Context Management component to support the analysis:

Table 1: User interactions collected from Generic Dashboard

| Root Asset Interactions | Widget Interactions |
| --- | --- |
| Consume Root Asset | Scroll Widget |
| Scroll Root Asset | Dismiss Widget (Close window) |
| Fullscreen Root Asset | Like Widget |
| Comment Root Asset | Dislike Widget |

### 3.2. SAM's Graph database

The Graph database of SAM is composed of edges and nodes, with nodes representing entities and edges relations between them. Three types of entities are defined: "Assets", "Persons" and "Keywords". Assets represent all kind of multimedia content in SAM (video, widgets, related information etc.), while Keywords are nodes describing the tags of an Asset. Finally, Persons represent users of the SAM first and second screen. Every type of node has specific attributes, describing the information it enfolds. For instance, an Asset has attributes such as id, type, title, etc. and a Person has name, identifier, etc.

Nodes are connected to each other with edges called relationships. Assets can be connected with other Assets with the relationship "is root asset of", signifying the widgets of a root Asset (movie). Assets can also be connected with Keywords with the relationship "has keywords" or with Persons via a variety of relations. For Root Assets (movies), these relationships are "consumes" and "comments". If a person has watched a movie or consumed related material then automatically an edge describing "consume" action is created to store this action. The same principle is applied to the "comment" relationship: if users express an opinion about an asset, the action is stored as "comment". Other relationships existing in such connections are "dislike", "like", "fullscreen" and "scroll". Edges have also attributes in order to enrich the information about the entities' relationships. For example the edge "comment" contains information about the intensity of a comment, its type (if it is a negative or a positive comment) and the comment itself. An instance of the SAM graph DB with some initial records can be seen in Figure 2:

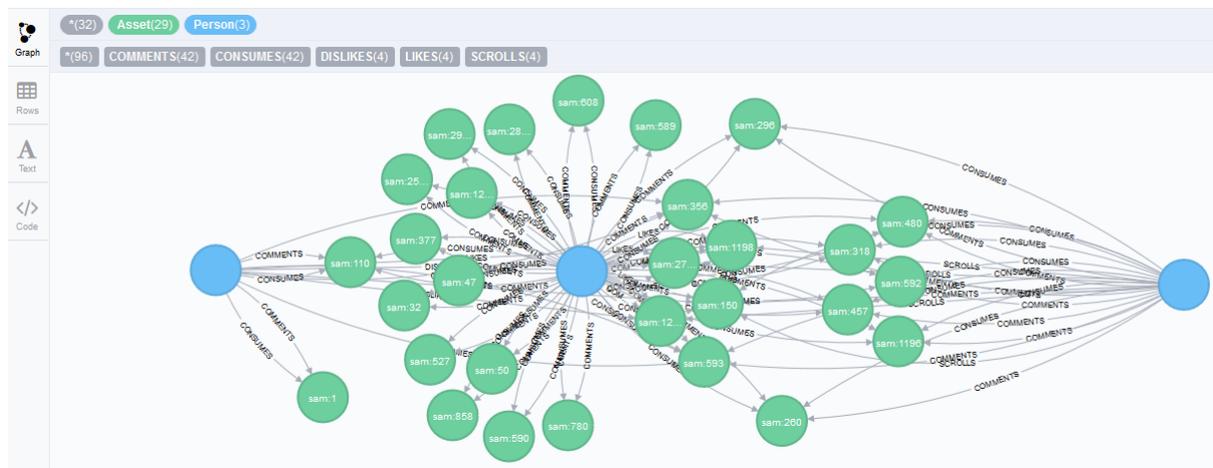

**Figure 2**: SAM Context Manager graph database example

Note that the same person can change her mind on an asset (e.g. dislike what she liked in the past). Thus, we decided to replace the explicit interactions (like/dislike, comment) with the latest one received from the SAM Dashboard, so that the relevance score to be calculated is efficient and properly updated.

## 4. Context Analysis and Recommendations

### 4.1. Graph analysis

A basic part of the graph analysis is the application of "weights" to the interactions between users and assets. Setting +1 and -1 as absolute values of relevancy and irrelevancy respectively, we apply those values to user-asset relations that explicitly show such a rating ("like" weights for +1, "dislike" weights for -1). Moreover, comments on assets are saved along with their sentiment polarity and intensity (percentage of positivity or negativity), thus we can apply decimal weights, ranging from (0,+1] for positive comments and from [-1,0) for negative comments. Zero values obviously express neutrality.

However, consuming or scrolling a root asset also indicates some interest by the user. The same applies for clicking or scrolling a specific widget in second screen, while dismissing it before it automatically closes indicates lack of interest. To capture those implicit patterns, we need to make sure that they will not totally overlap the explicit ones already mentioned. For example, if a user has "liked" an asset, but on the other hand dismissed it early on, this implies a weaker "like" or "interest" relation. The approach that we follow to make sure the overall weight (sum of weights) is mainly defined by "likes" / "dislikes" and only partly affected by other interactions is to apply to the latest a weight of

$$w_i = \frac{p_i}{t-1} \quad (1)$$

where $p_i$ = polarity indication(+1,-1) and $t$ = number of interaction types for this asset type. In this case, if an explicit interaction weight $w_e$ is contradictory to implicit weights $w_i$, the overall weight $W = w_e + \sum w_i$ will still bare the (now normalized) "polarity" of $w_e$.

Moreover, we want to collect indirect user relationships with an asset. In cases, for example, that a user has "liked" or commented positively for all widgets or keywords of a root asset (which may also exist in other videos as well), a strong indication of relevance to this root asset also exists. Similarly to the previous logic, we need to make sure that indirect relations to assets will not overlap a direct weight to it. Thus, for every rating to a connected asset/keyword we apply a weight of

$$w_x = \frac{r_x}{a+k+1} \quad (2)$$

where $r_x$ = rating of neighbouring node, $a$ = number of neighbouring assets and $k$ = number of the initial asset's keywords. The overall relevance weight of a person for an asset now becomes:

$$W = w_e + \sum w_i + \sum w_x \Leftrightarrow W = w_e + \sum \frac{p_i}{t-1} + \sum \frac{r_{xi}}{a+k+1} \tag{3}$$

Running this process recursively for connected assets, we conclude to the following general algorithm:

*Pseudo-code of Context Management graph analysis to rate the relevance of an asset for a user.*

```
procedure AssetRelevance (user, asset, graph)

  RelevanceWeights we, wi, wx;
  InteractionList I;
  InteractionTypes types= Interaction.getTypes();
  t = types.length();

  if user.hasRated(asset) then
    we = user.getExplicitRating (asset);

  if user.hasInteracted(asset) then
    I = user.getInteractions(asset);
    foreach interaction ∈ I do
      wi = wi + interaction.getPolarity()/(t-1);

  A=asset.getNeighbours();
  n = A.length();
  foreach neighbour_asset ∈ A do
     rx = AssetRelevance (user, neighbour_asset, (graph-asset));
     wx = wx + rx /(n+1);

return we + wi + wx;
```

Note that keywords are also treated as assets in the following algorithm for simplicity purposes.

To provide a rated list of assets to a user, based on her relevance to those, we need to calculate a user's node weights with existing assets. Thus, the algorithm shown above must be run for all assets in the graph which implies a complexity of $O(n^2)$. However, given the fact that recursive calls only need to apply for depth 2 in order to make sense (when shortest path between assets equals 2 to bare some meaningful relevance), complexity is further reduced to $O(n)$.

**4.2. Collaborative filtering analysis**

In cases of more "isolated" assets in the graph, when the user analyzed has not interacted with those or their neighbours (e.g. a new movie), it is obvious that the aforementioned analysis will not identify any meaningful relevance. In such cases, it was decided to use collaborative filtering among different users, in order to estimate the user relevance with the specific assets, based on her correlation with other users.

A most common approach used for collaborative filtering, having a dataset of simple numeric ratings[15], is using the Pearson Correlation Coefficient:

$$c_{au} = \frac{\sum_{i=1}^{h}(r_{ai} - \bar{r}_a) \times (r_{ui} - \bar{r}_u)}{\sqrt{\sum_{i=1}^{h}(r_{ai} - \bar{r}_a)^2 \times \sum_{i=1}^{h}(r_{ui} - \bar{r}_u)^2}} \quad (5)$$

between users *a* and *u*, where in our case $h = |I_{au}|$ is the amount of assets having been rated by both users, $r_{ai}$ is user *a*'s weight for asset *i* and $\bar{r}_a = average(r_{a1}, r_{a2}, \ldots, r_{ah})$.

Having calculated the correlation coefficients of a user with other users, collaborative filtering analysis can provide a prediction, rating her relevance with an asset *j*, based on other users' relevance for the specific asset and their correlation:

$$p_{aj} = \bar{r}_a + \frac{\sum_{u=1}^{g}(r_{uj} - \bar{r}_u) \times c_{au}}{\sum_{u=1}^{g} c_{au}} \quad (6)$$

where *g* is the number of users that consumed *j* and $p_{aj}$ is the predicted rating of relevance for user *a*.

**4.3. Personalised recommendations**

The results of the analysis processes described above for every user is two-fold:

- A rated list of root assets, consisted of pairs of videos and relevancy scores for the user, similar to top-k ranking approach presented in [16]
- A rated list of sub-assets of any root asset (appearing as widgets in second screen), consisted of pairs of widgets and relevancy scores for the user

Thus, personalized recommendations can be provided to first and second screen Syndicator component, to prioritize or even disappear irrelevant movies and related widgets of a movie upon consumption.

This results into the following two-level recommendation mechanism that provides:

- Smart recommendations of root assets (videos) to user
- Smart recommendations of second screen widgets to user, while watching a Smart TV program on first screen

## 5. Experiments and Evaluation

**5.1. Dataset and Configuration**

The presentation of the analysis methods above makes the performance advantages of the graph-based approach evident. For example, when retrieving for users having consumed a specific asset in the graph, Neo4j just returns the neighbours of the corresponding node, in contrast with the latency resulting from a respective SQL query. To acquire a meaningful and extended dataset, in order to form the SAM experimental Context Management graph, the authors decided to search online for available related data (users, movies, keywords, likes etc.).

Our experimental dataset is comprised of a big movie rating dataset found online [23], comprising a huge database of movies and user ratings, as well as keywords linked with those movies. To get most correlated users and movies in order to make a meaningful graph collaborative filtering and analysis, we selected the 30 most popular movies, in terms of number of ratings, rated by 619 users (overall 7038 ratings) and the 5 most popular keywords for each movie. This sampling assures that many common keywords exist between movies, so that the aforementioned graph weighting will apply. The dataset imported was interpreted into the SAM logic, directly importing SAM users and assets (movies and keywords) into the graph. Unfortunately the original dataset does not contain real life user interaction data, thus ratings were used to simulate comment sentiments and likes/dislikes based on their values, without any other interactions (scroll, fullscreen etc.) captured in the graph. With no widgets and relevant interactions existing, the current evaluation results are limited only to the first screen. The analysis in the context of the current experiments results into a movie relevance evaluation, which can then produce movie recommendations (one level evaluation) for users based on their movie relevance.

The dataset was split into training and testing sets (70% and 30% of the original rating respectively), with the first one to be fed into various data analysis algorithms and the later to be used as ground truth. Experiments conducted mostly focused on the accuracy of the aforementioned technique, in contrast with mainstream machine learning approaches, for estimating users' relevance with movies of the testing set. The graph analysis, supported by the Pearson Collaborative filtering presented above, was applied and compared with a k-nearest neighbours (K-NN) algorithm run over Neo4j[4], as well as various machine learning algorithms (SVM, C4.5, MLP etc.) employed in Weka software, version 3.7[5], taking the initial dataset as an .arff file input. Experiments operated on a desktop machine with Intel Core TM i5-3400 Processor 2.80 GHz and 12GB of RAM memory, running 64-bit Windows 10 Pro.

**5.2. Experimental Results**

In Table 2 an analytical report of results per approach is provided, for predicting users' relevance to the assets (movies) of the training set. The mean errors refer to the difference between the calculated values and the real original ratings, ranging from -1 to 1, which represent the current evaluation's ground truth.

Note that time comparison between algorithms running in batch mode that do not connect to a database and algorithms that return results on-demand, like the one we implement in SAM, is irrelevant. For example the Naive Bayes approach trains its model instantly but requires the full dataset available in memory. When considering our dataset of 7038 ratings this is feasible, however it is apparent that this is not a scalable solution. The motivation behind the decision to get metrics from state-of-the-art machine learning algorithms is mainly to evaluate the accuracy of our proposed algorithm.

---

[4] https://neo4j.com/graphgist/8173017/
[5] http://www.cs.waikato.ac.nz/ml/weka/

Table 2: Accuracy and mean errors of SAM predictions, compared to State-of-the-Art machine learning techniques

|  | Mean absolute error | Root mean squared error | Mean percentage error |
|---|---|---|---|
| **SVM with linear kernel** | 0.1099 | 0.3315 | 5.5% |
| **C4.5 w/ 10 Bagging** | 0.1253 | 0.2688 | 6.3% |
| **Best-first decision tree** | 0.1274 | 0.2626 | 6.4% |
| **Logistic Regression** | 0.1269 | 0.2612 | 6.4% |
| **LAC Lazy Associative Classifier** | 0.1306 | 0.2563 | 6.5% |
| **Bayes Net with K2 search** | 0.1263 | 0.2559 | 6.3% |
| **NaiveBayes** | 0.1283 | 0.2551 | 6.4% |
| **Naive Bayes Tree** | 0.1334 | 0.2652 | 6.7% |
| **MLP 100 neurons** | 0.1324 | 0.2632 | 6.6% |
| **CNN 100 neurons** | 0.1269 | 0.2530 | 6.4% |
| **CNN 1000 neurons** | 0.1202 | 0.2578 | 6.0% |
| **Hoeffding Tree** | 0.1499 | 0.2684 | 7.5% |
| **Hidden Markov Model** | 0.1653 | 0.2875 | 8,3% |
| **K-nearest neighbour on Neo4j** | 0.3226 | 0.4108 | 16,1% |
| **SAM (graph + CollabFiltering)** | **0.1312** | **0.2604** | **6.6%** |

As can be observed in the experimental results, SAM's approach scores relatively well, taking into account the low performance cost introduced by the graph database, as well as its multilevel capabilities. The authors compared our graph-based analysis approach with variations of some known algorithms. In particular, we used Breiman's Bagging technique [24] with 10 iterations (C4.5) and also tried different neural network models (Convolutional Neural Network with 100 and 1000 neurons, Multi-layer Perceptron). Probabilistic models (Logistic Regression, Naive Bayes, Bayesian network with K2 search algorithm) and simple tree implementations (C4.5, Best-first) were also used, yielding competitive results. In terms of mean absolute error the SVM with a linear kernel was superior but with a weak root mean squared error, for which the 100 neurons CNN produced the best results. We approached the problem as a classification problem because the dataset has discrete values in ratings (1-5); however we report the error metrics because in our case we are interested in how close the predictions were to the ground truth. Since our data is ordinal and have an inherent order, it is a safer metric to ensure generalization of models.

Beyond comparing SAM Context Management approach with other machine learning techniques in terms of effectiveness, we also present a comparison of the two algorithms running on top of the graph database (SAM, K-NN) time-wise. Resulting from those algorithms, SAM's Context Management component exposed two recommendation web services respectively. Stress-testing our approach performance, as a commercial deployed service, we used JMeter [25] to generate requests for all ratings of the testing set to those services. The response times of the first 1000 requests can be seen in Figure 3:

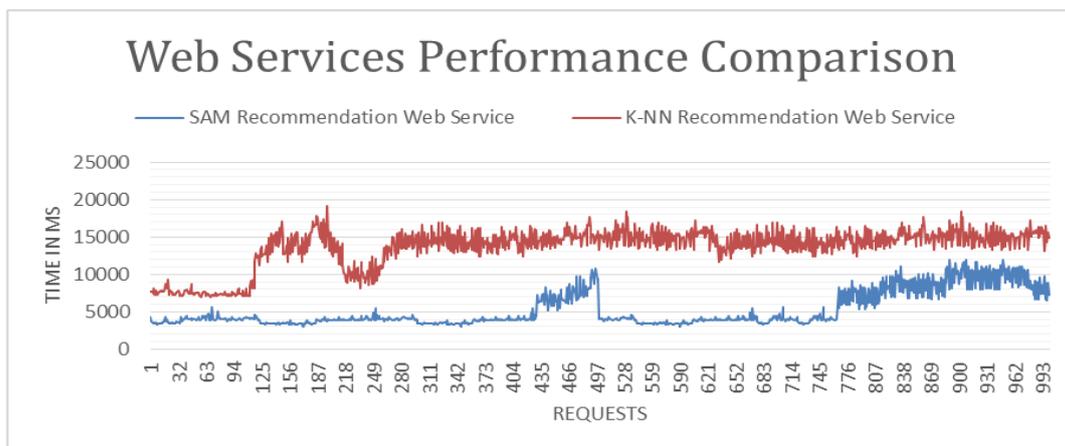

Figure 3: A time performance comparison between K-NN and SAM Recommendation Web Services

SAM Recommendation Web Service evidently outperforms K-NN, one of the most popular clustering approaches for recommendations using graphs. Results above validate the current model's low complexity and illustrate its scalability. Results can be explained by the locality of the data needed for SAM. In both recommendations levels the algorithms start from the user's node and travel in his near neighbours to find interactions without the need of prior knowledge that is needed in the K-NN case.

## 6. Conclusions

In this paper, we presented an efficient Context Management approach for Social TV users, collecting context-related data and actions to provide personalized multi-level recommendations via a hybrid method combining graph paths analysis and Pearson collaborative filtering. Experiments used a real movie rating dataset found online and illustrated promising results in one level recommendations, in terms of accuracy and performance. The effectiveness of the current model will be more evident with the addition of an extended user interaction dataset, which can be aggregated from the SAM second screen listeners during the final trials of the project in the upcoming months. Thus, in the future the authors plan to aggregate Social TV-related datasets, in order to evaluate the current model end to end in both levels (first and second screen). The validity of the results will be better illustrated using more diverse datasets, in terms of user relevance scores, thus training sets where ratings are distributed more widely must be included, so that correlation techniques can generate more concrete and meaningful user clusters. Lastly, as long as such a solution will go commercial, privacy-related issues should be also tackled, employing user anonymisation and privacy-preserving item-based collaborative filtering [26].

## 7. Acknowledgment

This work has been supported by the SAM project and funded from the European Union's 7th Framework Programme for research, technological development and demonstration under grant agreement no 611312.